\documentstyle[aps,pra,multicol,amssymb,epsf]{revtex}

\title{Decoherence in nonclassical motional states of a trapped ion}

\author{M. Murao and P.L. Knight}

\address{Optics Section, Blackett Laboratory, Imperial College, 
London SW7 2BZ, UK}

\date{\today}

\begin{document}
\draft
\maketitle
\begin{abstract}
The decoherence of nonclassical motional states of a trapped $^9 {\rm
Be^+}$ ion in a recent experiment is investigated theoretically.
Sources of decoherence considered here destroy the characteristic
coherent quantum dynamics of the system but do not cause energy
dissipation. Here they are first introduced phenomenologically and
then described using a microscopic Hamiltonian
formulation. Theoretical predictions are compared to experimental
results.
\end{abstract}


\begin{multicols}{2}

\section{Introduction}

The experiments of Meekhof et al \cite{Meekhof} have revealed quantum
dynamics characteristic of the Jaynes-Cummings type (especially
collapses and revivals of excitation probabilities) \cite{Shore} for
the first time in a trapped ion system.  Stimulated Raman transitions
coupled the internal states of a trapped $^9 {\rm Be^+}$ ion to its
motional states, within the Lamb-Dicke limit of tight ion motion
confinement in the trapping potential.  The Jaynes-Cummings spin-boson
Hamiltonian then derives from the coupling of the internal electronic
states of the ion to the vibrational quantum states of motion.

The characteristic quantum dynamics (collapse and revival) of the
Jaynes-Cummings type interaction for the ion motion
\cite{Blockley,Leibfreid,Wineland} (in the experiment of Meekhof et al
\cite{Meekhof}, an ``{\it anti} Jaynes-Cummings interaction'' for
driving the first blue sideband) were observed in the population of
the lower atomic state ($P_\downarrow$), which was modelled by the
phenomenological form fitting the observation as
\begin{eqnarray}
P_\downarrow \left( t \right) = \frac{1}{2} \left \{ 1+ \sum_{n} {p_n
\cos \left( 2 g t \sqrt{n+1} \right) {\rm e}^{-\gamma_n t} } \right
\}.
\label{eqn:probexper}
\end{eqnarray}
Here $p_n$ is the initial probability distribution for the motional
states in the Fock state basis, $g$ is a coupling constant between the
motional states and atomic states (Rabi frequency), and $\gamma_n$ is
a phenomenological damping rate.  The observed damping rate can be
written as $\gamma_n=\gamma_0 (n+1)^\nu$ with $\nu \approx 0.7$
observed in the experiments of Ref \cite{Meekhof}.  The damping rate
of the $n$th component is independent of that of different components,
so that equation (\ref{eqn:probexper}) implies decoherence without
there being transitions between the states of different quantum
numbers (energy relaxation).  The conventional sources of decoherence,
such as spontaneous emission between internal atomic states, and
population decay of motional states, cause transitions between the
states of different quantum numbers and do not give the decay rate in
a form which can be written as $\gamma_n$.  There have been
suggestions \cite{Wineland} as to the origin of this decoherence with
the unusual observed value of $\nu$, in terms of decoherence of the
ion motion, decoherence of the ion internal levels, and decoherence
caused by non-ideal applied fields but the situation has not yet been
satisfactorily resolved.

In this paper, we introduce phenomenologically new sources of
decoherence, which destroy the characteristic Jaynes-Cummings type
dynamics without energy relaxation, by coupling the spin-boson system
to a quantum reservoir \cite{Bose}.  The reservoir consists of
many-mode bosons described by a canonical distribution at temperature
$T$ and introduces noise to the system.  We treat decoherence
microscopically using a master equation.  The master equation
coincides with that for stochastic white noise in the high temperature
limit of the reservoir under certain approximations (Markovian
approximation and ohmic density of states of the reservoir
\cite{Leggett}).  The advantage of using a quantum boson reservoir is
that it not only describes phenomenological quantum noise, but also
gives more microscopic information on the source of decoherence,
e.g. the noise frequency being responsible for decoherence even in the
high temperature limit.  Using this combined approach from two
directions (phenomenological and microscopic) we discuss the origins
of decoherence in this system.

\section{Description of the system without decoherence}

Before investigating decoherence, we consider the system without
decoherence, reviewing how the stimulated Raman transitions describe
the ``anti Jaynes-Cummings'' interaction \cite{Blockley,Leibfreid,Wineland}
when the first blue sideband is driven, and introducing the dressed
states description of the anti Jaynes-Cummings system.  We note that
the first red sideband driven case (the Jaynes-Cummings interaction
case) can be treated just in the same manner, where we exchange the
two relevant internal atomic levels $\left \vert \downarrow \right
\rangle$ and $\left \vert \uparrow \right \rangle$ of the following
formulation.

We consider a system with three internal levels $\left \vert j \right
\rangle$ ($j=\underline{0},\downarrow,\uparrow$) and their motional
states $\left \vert n \right \rangle$ ($n=0,1,....$).  They are
represented by the following Hamiltonian:
\begin{eqnarray}
H_0=H_{atom}+H_{vib}
\end{eqnarray}
where
\begin{eqnarray}
H_{atom}&=&
-\hbar \omega_{01} \vert {\downarrow} \rangle \langle {\downarrow} \vert
-\hbar \omega_{02} \vert {\uparrow} \rangle \langle {\uparrow} \vert, \\
H_{vib}&=&\hbar \omega_x b^\dagger b
\end{eqnarray}
with the transition frequency $\omega_{01}$ ($\omega_{02}$) between
states $\left \vert \downarrow \right \rangle$ ($\left \vert \uparrow
\right \rangle$) and $\left \vert \underline{0} \right \rangle$, the
creation (annihilation) operator of the motional states $b^\dagger$
($b$), and the frequency of the motional states $\omega_x$.  We employ
two driving laser beams with detuning $\Delta$, momentum ${\bf k_1}$
(${\bf k_2}$) and frequency $\omega_1$ ($\omega_2$) which cause dipole
transitions between the level $\left \vert \downarrow \right \rangle$
($\left \vert \uparrow \right \rangle$) and $\left \vert \underline{0}
\right \rangle$.  (See Fig.~\ref{fig:ionlevel}) These beams can be
treated classically, so the interaction Hamiltonian is
\begin{eqnarray}
H_{int}= -\mbox{\boldmath $\mu$}_{01} \cdot {\rm Re} \left [ {\bf
E}_{01} {\rm e}^{i \left({\bf k}_1 \cdot {\bf r}-\omega_1 t \right)}
\right ] \nonumber \\ -\mbox{\boldmath $\mu$}_{02} \cdot {\rm Re}
\left [ {\bf E}_{02} {\rm e}^{i \left({\bf k}_2 \cdot {\bf r}-\omega_2
t \right)} \right ]
\label{eqn:int0}
\end{eqnarray}
where $\mbox{\boldmath $\mu$}_{0 1}$ ($\mbox{\boldmath $\mu$}_{0
2}$) is the dipole matrix element between $\left \vert \downarrow
\right \rangle$ ($\left \vert \uparrow \right \rangle$) and $\left
\vert \underline{0} \right \rangle$.

\begin{minipage}{3.27truein}
\begin{center}
\begin{figure}
  \begin{center}
    \leavevmode
    \epsfxsize=3.truein
    \epsfbox{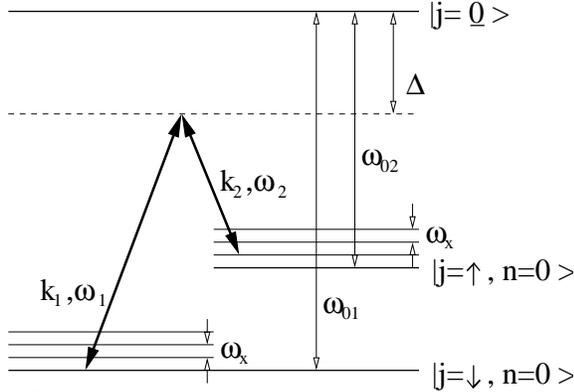}
  \end{center}
\caption{Energy levels of the internal states and the motional
states.}
\label{fig:ionlevel}
\end{figure}
\smallskip
\end{center}
\end{minipage}

We apply the rotating wave approximation to the interaction
Hamiltonian (\ref{eqn:int0}), transform to the interaction picture
($H_{int}^I={\rm e}^{i H_0 t/\hbar} H_{int} {\rm e}^{-i H_0 t/\hbar}
$) and expand in terms of the motional state quantum numbers. When the
blue sideband is driven, we have
\begin{eqnarray}
H_{int}^I &=&
-\hbar \sum_{n,m}{
g_{01} {\rm e}^{i \left\{
\left(n-m \right) \omega_x + \mit\Delta \right \} t}
\left\langle {n} \right\vert
{\rm e}^{i {\bf k}_1 \cdot {\bf r}}
\left\vert {m} \right\rangle
\left\vert {n} \right \rangle
\left\langle {m} \right \vert
\sigma_1^+} 
\nonumber \\
&-& \hbar
\sum_{n,m}{
g_{02} {\rm e}^{i \left\{
\left(n-m+1 \right) \omega_x + \mit\Delta \right \} t}
\left\langle {n} \right\vert
{\rm e}^{i {\bf k}_2 \cdot {\bf r}}
\left\vert {m} \right\rangle
\left\vert {n} \right\rangle
\left\langle {m} \right\vert
\sigma_2^+}\nonumber \\
&-& h.c.
\end{eqnarray}
where we have introduced the dipole operators $\sigma_1^+ \equiv
\left\vert {\underline{0}} \right \rangle \left\langle {\downarrow}
\right \vert$, $\sigma_2^+ \equiv \left\vert {\underline{0}} \right
\rangle \left\langle {\uparrow} \right \vert$, and the quantities
$g_{01} \equiv \mbox{\boldmath $\mu$}_{01} \cdot {\bf E}_{01}/2
\hbar$, $g_{02} \equiv \mbox{\boldmath $\mu$}_{02} \cdot {\bf
E}_{02}/2 \hbar$.

The large detuning condition allows the adiabatic elimination of the
level $\left \vert \underline{0} \right \rangle$ \cite{Steinbach}.
Under this condition, Raman transitions dominate the system.  We also
assume the system is cool enough to reach the Lamb-Dicke limit ($ {\bf
k}_{j^\prime} \cdot {\bf r} \ll 1$) so we can expand
\begin{eqnarray}
{\rm e}^{i {\bf k}_{j^\prime} \cdot {\bf r}} \approx 1+i k_{j^\prime
x} x_0 \left(b+b^\dagger \right),
\end{eqnarray}
where $j^\prime=1,2$, $x_0 \equiv \left( \hbar/2m \omega_x
\right)^{1/2}$ and $m$ is the mass of the ion. Then the effective
Hamiltonian in the interaction picture can be written
\begin{eqnarray}
H_{int}^I= \hbar \Delta_1 
\left \vert \downarrow \right \rangle
\left \langle \downarrow \right \vert 
+ \hbar \Delta_2
\left \vert \uparrow \right \rangle
\left \langle \uparrow \right \vert \nonumber \\
+ 
\hbar g b^\dagger 
\left \vert \uparrow \right \rangle
\left \langle \downarrow \right \vert
+
\hbar g^* b 
\left \vert \downarrow \right \rangle
\left \langle \uparrow \right \vert
\end{eqnarray}
where $\Delta_l \equiv \left \vert g_{0l}\right \vert^2 / \mit\Delta$
and $g \equiv i g_{01}^* g_{02} \delta k_x x_0 / \mit\Delta$ with
$\delta k_x=k_{2x}-k_{1x}$.  If we write $S_+ \equiv \left \vert
\uparrow \right \rangle \left \langle \downarrow \right \vert$, $S_-
\equiv \left \vert \downarrow \right \rangle \left \langle \uparrow
\right \vert$, remove the terms for energy shifts and set $g=g^*$, the
effective Hamiltonian has the anti Jaynes-Cummings form
\begin{eqnarray}
H_{eff}^{I}=\hbar g \left( b^\dagger S_+ + b S_- \right).
\label{eqn:hinteff}
\end{eqnarray}
This effective Hamiltonian (\ref{eqn:hinteff}) is the origin of the
characteristic quantum dynamics (Rabi oscillations, collapses and
revivals) of the system.  Decoherence is the decay of the off-diagonal
elements which represent the characteristic quantum dynamics, so we
can use this Hamiltonian to explore some of the sources of decoherence
in this interaction picture in the next section.

When working in the interaction picture, it is convenient to
introduce the dressed states for the effective Hamiltonian
(\ref{eqn:hinteff}):
\begin{eqnarray}
&&\left \vert {\varphi\left(n,1\right)} \right \rangle
=\frac{1}{\sqrt{2}} \left( \left \vert {\downarrow,n} \right \rangle
+\left \vert {\uparrow,n+1} \right \rangle \right),
\label{eqn:eigens1}
\\
&&\left \vert {\varphi\left(n,2\right)} \right \rangle
=\frac{1}{\sqrt{2}} \left( \left \vert {\downarrow,n} \right \rangle 
-\left \vert {\uparrow,n+1} \right \rangle
\right) 
\label{eqn:eigens2}
\\
&&\left \vert {\uparrow,0} \right \rangle, 
\label{eqn:eigens0}
\end{eqnarray}
which are the eigenstates of the effective Hamiltonian.   We write the
eigenvalue of (\ref{eqn:eigens1}) as $E_+^n$, of (\ref{eqn:eigens2})
as $E_-^n$, of (\ref{eqn:eigens0}) as $E_0$, so we have $E_\pm^n=\pm
\hbar g \sqrt{n+1}$ and $E_0=0$.  We write the reduced density
operator in the dressed state basis as
\begin{eqnarray}
\rho^I \left( t \right)
&=& \sum_{n,\alpha} {\sum_{m,\beta}{\rho_{\alpha \beta}^{n m}\left(t\right)}}
\left\vert {\varphi \left(m,\beta \right)} \right\rangle 
\left\langle {\varphi \left(n,\alpha \right)} \right\vert 
\nonumber\\
&+& \sum_{n,\alpha}{\rho_{\alpha 0}^{n}\left(t\right)}
\left\vert {\uparrow,0} \right\rangle
\left\langle {\varphi\left(n,\alpha \right)} \right\vert 
\nonumber \\
&+& \sum_{m,\beta}{\rho_{0 \beta}^{~m} \left(t\right)}
\left\vert {\varphi\left(m,\beta \right)} \right\rangle 
\left\langle {\uparrow,0} \right\vert 
\nonumber\\
&+&\rho_{00}\left(t\right)
\left\vert {\uparrow,0} \right\rangle
\left\langle {\uparrow,0} \right\vert 
\end{eqnarray}
for the boson quantum numbers $n, m=0,1,2,...$ and the spin quantum
numbers $\alpha, \beta=1,2$, where $\rho_{\alpha \beta}^{nm}$,
$\rho_{\alpha 0}^{n}$,$\rho_{0\beta}^{~m}$, $\rho_{00}$ are matrix
elements.  Then the population of the lower atomic state
$P_\downarrow$ is
\begin{eqnarray}
P_\downarrow \left( t \right)
= \frac{1}{2} \left(1-\rho_{00} \left( t \right)
+2 \sum_n{{\rm Re} \left[ \rho^{nn}_{12} \left( t \right) \right]} 
\right). 
\end{eqnarray}
Note that only the elements that are off-diagonal in terms of the spin
quantum number ($\rho^{nn}_{12} \left( t \right)$) and one diagonal
element ($\rho_{00} \left( t \right)$) contribute to $P_\downarrow$ in
the dressed state basis.  Basically, the characteristic quantum
dynamics observable in the population of the lower state are due to
the dynamics of elements that are off-diagonal in terms of the spin
quantum number.

\section{Decoherence without energy relaxation}

We next consider the system with the effective Hamiltonian
(\ref{eqn:hinteff}) described in the previous section now surrounded
by the environment, that is, as an open system.  Noise from the
environment causes decoherence \cite{Louisell,Zurek}.  We treat this
open system by coupling to a quantum reservoir, which consists of an
infinite number of many mode bosons
\begin{eqnarray}
H_{r}=\hbar \sum_{l} \omega_l B_l^\dagger B_l,
\end{eqnarray}
where $\omega_l$ is the $l$th reservoir frequency, $B_l^\dagger$ and
$B_l$ are the creation and annihilation operators of the reservoir
bosons.  Since the reservoir has infinitely greater degrees of
freedom, the reservoir bosons are not affected by the system.  Then
the time evolution of the reservoir boson operators are given by
\begin{eqnarray}
B_l^\dagger \left( t \right) 
={\rm e}^{i H_r t/\hbar} B_l^\dagger {\rm e}^{-i H_r t/\hbar}
={\rm e}^{i \omega_l t} B_l^\dagger ,\\
B_l \left( t \right) 
={\rm e}^{i H_r t/\hbar} B_l {\rm e}^{-i H_r t/\hbar}
={\rm e}^{-i \omega_l t} B_l.
\end{eqnarray}
The system-reservoir coupling Hamiltonian is
\begin{eqnarray}
H_{sr}=\hbar \sum_s{C_s} \sum_l{g_{sl} \left( B_l^\dagger+B_l\right)}
\label{eqn:srint}
\end{eqnarray}
where $g_{sl}$ is the coupling between a system operator $C_s$ and the
$l$th reservoir mode.  The sum of the system operators $\sum_s{C_s}$
has to be Hermitian. In the master equation derived from the
system-reservoir coupling (\ref{eqn:srint}), the damping term consists
of the system operators coupling to the reservoir operators.  Thus the
choice of the coupling between system operators and the reservoir
determines the effect of the reservoir.  If we choose a system
operator $C_s$ with the property
\begin{eqnarray}
C_s \left \vert {\varphi\left(n,\alpha \right)} \right \rangle =
\sum_\beta {c_\beta} \left \vert 
{\varphi\left(n,\beta \right)} \right \rangle
\end{eqnarray}
the resulting master equation describes relaxation within the dressed
states of the quantum number $n$, but not energy relaxation between
states with different $n$.  This is because the time evolution of the
density matrix elements in terms of $\left \vert
{\varphi\left(n,\alpha \right)} \right \rangle$ decouples for
different $n$.  The operators $S_z$, $b^\dagger b$ are obviously of
this type, as these operators do not even change the motional states
$\left \vert n \right \rangle$ as well as the dressed state label $n$.
The operator $b^\dagger S_+ + bS_-$ changes the motional state, but
this operator does not change the dressed state occupation label $n$,
so $b^\dagger S_+ + b S_-$ is of this type, too. On the other hand, if
we choose $C_s$ with
\begin{eqnarray}
C_s \left \vert {\varphi\left(n,\alpha \right)} \right \rangle =
\sum_\beta {c_\beta^\prime} \left \vert {\varphi
\left(m \neq n,\beta \right)}
\right \rangle,
\end{eqnarray}
then the resulting master equation describes transitions between
states with different boson quantum numbers, which cause energy
relaxation;  $S_+ + S_-$ and $b+b^\dagger$ are of this type.

\subsection{Imperfect dipole transition}

First, we treat the case when the system operator which is coupling to
the reservoir is $b^\dagger S_+ + b S_-$.  This case looks strange at
first sight, but we can consider this as the result of imperfect
dipole transitions between the level $\left \vert \underline{0} \right
\rangle$ and the level $\left \vert j \right \rangle$
($j=\downarrow,\uparrow$) due to fluctuations of the driving laser
intensity. We have previously described how phase fluctuations lead to
decoherence and the destruction of quantum revivals in the
Jaynes-Cummings model \cite{Moya-Cessa}.  This is one particular
realization of ``intrinsic decoherence'' in which off-diagonal density
matrix elements relax {\it without} energy relaxation
\cite{Moya-Cessa,Milburn}.  We note that these earlier results of ours
apply to the experiments of Ref.~\cite{Meekhof} if the source of
decoherence is relative phase fluctuations driving the ionic Raman
transition.  Here we analyse more general sources of decoherence.  The
imperfect dipole transitions $\left \vert \downarrow \right \rangle
\Leftrightarrow \left \vert \underline{0} \right \rangle$ and $\left
\vert \uparrow \right \rangle \Leftrightarrow \left \vert
\underline{0} \right \rangle$ are represented by
\begin{eqnarray}
\sigma_1^\pm &\rightarrow& \sigma_1^\pm +\sigma_1^\pm
\sum_l{g_{l} \left( B_l^\dagger+B_l\right)},\\
\sigma_2^\pm &\rightarrow& \sigma_2^\pm +\sigma_2^\pm
\sum_l{g_{l} \left( B_l^\dagger+B_l\right)}.
\end{eqnarray}
We assume the system-reservoir coupling is weak enough, so we can
neglect the terms that are second order in $g_l$.  Then we have, for
example,
\begin{eqnarray}
S_+=\sigma_2^-\sigma_1^+ &\rightarrow&
\sigma_2^-\sigma_1^+ +\sigma_2^-\sigma_1^+
\sum_l{g_{l} \left( B_l^\dagger+B_l\right)}.
\end{eqnarray}
Thus the Hamiltonian describing the system-reservoir coupling is given
by
\begin{eqnarray}
H_{sr}&=&\hbar \left( b^\dagger S_+ +b S_-\right) \sum_l{g_{l}^\prime
\left( B_l^\dagger+B_l\right)}
\label{eqn:coupleSb}
\end{eqnarray}
where $g_l^\prime=g g_l$.  This Hamiltonian (\ref{eqn:coupleSb}) can
be interpreted as if the Rabi frequency ($g$) of the Jaynes-Cummings
type system fluctuates due to the system-reservoir coupling as
\begin{eqnarray}
g \rightarrow g+\sum_l{{g_l}^\prime \left( B_l+B_l^\dagger \right)}.
\end{eqnarray}

Using a time convolution-less (TCL) formulation (this approach is
described in detail in \cite{Shibata}) of the quantum damping theory
\cite{Louisell} and the rotating wave approximation on the master
equation \cite{Murao2} the master equation in the interaction picture
is
\begin{eqnarray}
    \frac {\partial}{\partial t}\rho^I \left(t\right)=
    \frac {1}{i\hbar}
    \left[ {H_{eff}^I,\rho^I \left( t \right)}\right]
     +{\it \Gamma} \rho^I \left( t \right) 
\label{eqn:mastereq1}
\end{eqnarray}
where the damping term ${\it \Gamma} \rho^I$ is given by \cite{Murao3} 
\end{multicols}
\begin{eqnarray}
{\it \Gamma} \rho^I \left( t \right) 
&=& \sum_l{{g_{l}^\prime}^2 \int_0^t{dt' \biggl\{\left( \langle
B_l^\dagger\left(t'\right)B_l\rangle_B+
\langle B_l\left(t'\right)B_l^\dagger\rangle_B \right) \biggr. }}
\nonumber\\
&\times& \left(\left[\underline{b^\dagger S_+}\left(-t'\right) 
\rho^I\left( t \right),b S_- \right]+
\left[\underline{b S_-}\left(-t'\right) \rho^I\left( t \right),
b^\dagger S_+ \right]
\right) \nonumber\\
&+&\left( \langle B_l^\dagger\left(-t'\right)B_l \rangle_B+
\langle B_l\left(-t'\right) B_l^\dagger \rangle_B \right) 
\nonumber\\
&\times& \biggl. \left(\left[b^\dagger S_+,
\rho^I\left( t \right)\underline{b S_-}\left(-t'\right)\right]+
\left[b S_-,
\rho^I\left(t\right)\underline{b^\dagger S_+}\left(-t'\right) \right]
\right)\biggr\} 
\label{eqn:dampint}
\end{eqnarray}
\begin{multicols}{2}
with
\begin{eqnarray}
\underline{b^\dagger S_+}(t)
= {\rm e}^{i H_{eff}^{I} t / \hbar} b^\dagger S_+ 
{\rm e}^{-i H_{eff}^{I} t / \hbar},\\
\underline{b S_-}(t)
= {\rm e}^{i H_{eff}^{I} t / \hbar} b S_- 
{\rm e}^{-i H_{eff}^{I} t / \hbar}.
\end{eqnarray}

The master equation (\ref{eqn:mastereq1}) can be solved by expanding
all system operators in terms of the dressed states under certain
reservoir conditions.  We require the reservoir to be the canonical
distribution at temperature $T$ and the time scale of the reservoir
variables to be much shorter than the system variables so we can take
the Markovian limit.

We take the continuum limit of the reservoir modes,
\begin{eqnarray}
\sum_l \rightarrow \int{d\omega D(\omega)}
\end{eqnarray}
where $D(\omega)$ is the density of states of the reservoir. The
corresponding continuum expression for $g_l$ is $g(\omega)$.  The
master equation is cast into a group of differential equations for the
density matrix elements.  The time evolution of density matrix
elements having different boson quantum numbers are decoupled due to
the character of the coupling between the system operator and the
reservoir.  The time evolution of the diagonal elements ($\rho_{\alpha
\alpha}^{nn}$) and the off-diagonal elements ($\rho_{\alpha
\beta}^{nn}$, $\alpha \neq \beta$) having the same boson number ($n$)
are also decoupled.

To calculate the time evolution of $P_\downarrow$, we only need the
elements $\rho_{12}^{nn}$ and $\rho_{00}$.  The equations for the time
evolution of these elements are 
\begin{eqnarray}
\frac {\partial}{\partial t}\rho_{00}\left(t\right)&=&0,\\
\frac {\partial}{\partial t}\rho_{12}^{nn}\left(t\right)&=&
- i \Omega_n \rho_{12}^{nn}\left(t\right) \nonumber \\
&-& \left( n+1 \right)
\left\{ \bar{n} \left( n \right)+1/2 \right\} \kappa \left(n \right)
\rho_{12}^{nn}\left(t\right) \nonumber \\ 
&-& 2 \left( n+1 \right) \kappa_0 \bar{n}_0 
\rho_{12}^{nn}\left(t\right) \nonumber \\
&-&  \left( n+1 \right) 
\left\{ \bar{n} \left( n \right)+1/2 \right\} \kappa \left(n \right)
\rho_{21}^{nn}\left(t\right), 
\label{eqn:diffequation}
\end{eqnarray}
where
\begin{eqnarray}
\Omega_n=E_n^+ - E_n^-=2 g \sqrt{n+1}.
\end{eqnarray}
In (\ref{eqn:diffequation}), the function of the reservoir bosons
$\bar{n}\left( n \right)$ is
\begin{eqnarray}
\bar{n} \left( n \right) =\left( {\rm e}^
{\hbar \Omega_n/k_B T} -1 \right)^{-1},
\label{eqn:bosonfunction}
\end{eqnarray}
and the damping function $\kappa \left( n \right)$ is
\begin{eqnarray}
\kappa \left( n \right) \sim D(\Omega_n) \cdot g(\Omega_n),
\label{eqn:dampfunct}
\end{eqnarray}
which represents the effective contribution of the reservoir bosons
having frequency $\Omega_n$.  So the combination of these ($\kappa
\left( n \right) \left \{ \bar{n} \left( n \right)+1/2 \right \}$)
represents the effective mean number of the reservoir bosons with
frequency $\Omega_n$.  The quantity $\kappa_0 \bar{n}_0$ is the
contribution from zero frequency reservoir bosons.

The analytical solution of the equation (\ref{eqn:diffequation}) is
\begin{eqnarray}
\rho_{12}^{nn}\left(t\right)
&=&{\rm e}^{-A_n t}
\left\{ \cos \left(B_n t \right)
-i {\Omega_n}/{B_n} \sin \left( B_n t \right)
\right \} \rho_{12}^{nn}\left(0 \right) \nonumber \\ 
&-&{\rm e}^{-A_n t} {A_n}/{B_n} \sin \left( B_n t \right)
\rho_{21}^{nn}\left(0 \right)
\label{eqn:analytical}
\end{eqnarray}
where
\begin{eqnarray}
A_n&=& \left(n+1 \right) \kappa \left( n \right)
\left \{ \bar{n} \left( n \right)+1/2 \right \}
+2 \left( n+1 \right) \kappa_0 \bar{n}_0 \nonumber \\
&\equiv& A_n^{dipole}, 
\label{eqn:damprate} 
\\
B_n&=&\sqrt{\Omega_n^2-A_n^2}.
\label{eqn:coherentmotion}
\end{eqnarray}
With the chosen initial condition:
$\vert \downarrow \rangle \langle \downarrow \vert$ for the atom and
$\sum_n p_n \vert n \rangle \langle n \vert$ for the motional state,
the real part of Eq.~(\ref{eqn:analytical}) is found to be
\begin{eqnarray}
{\rm Re}\left[ \rho_{12}^{nn}\left(t\right) \right]= 
\frac{{\rm e}^{-A_n t}}{2} \sqrt{1+\left(\frac{A_n}{B_n} \right)^2}
\cos \left(B_n t+\theta_n \right)
\label{eqn:realpart}
\end{eqnarray}
where $\theta_n$ is a phase shift defined by $\theta_n=\arctan
\left( A_n/B_n \right)$.  Thus we see that the damping rate $A_n$
depends on the effective mean number of the reservoir bosons with
frequency $\Omega_n$, with a factor $n+1$.  The
coupling to the reservoir also shifts the oscillation from $\Omega_n
t$ to $B_n t+\theta_n$.  Since we assumed that the system-reservoir
coupling is weak in our formulation, $\kappa \left(n \right)$ in $A_n$
must be much smaller than the Rabi frequency $g$.  Thus we have
relation $A_n \ll B_n \sim \Omega_n$.  Under this condition, the
population of the lower atomic state $P_\downarrow$ is approximated to
be
\begin{eqnarray}
P_\downarrow \left( t \right)
= \frac{1}{2} \left \{ 1
+\sum_n{p_n 
\cos \left(\Omega_n t \right) 
{\rm e}^{-A_n t}} \right \},
\label{eqn:pdownthoery}
\end{eqnarray}
which is in the same form as that seen in the experiments
\cite{Meekhof}.

\subsection{Fluctuation of vibrational potential}

Next, we consider the case that the system couples to the reservoir
via the system operator $b^\dagger b$.  The system-reservoir coupling
Hamiltonian is
\begin{eqnarray}
H_{sr}=\hbar b^\dagger b \sum_l{g_{l}^\prime \left(
B_l^\dagger+B_l\right)}.
\end{eqnarray}
This coupling describes fluctuations of the trap potential. Then the
damping term is
\begin{eqnarray}
{\it \Gamma} \rho^I \left( t \right) 
&=& \sum_l{{g_{l}^\prime}^2 \int_0^t{dt' \biggl\{\left( \langle
B_l^\dagger\left(t'\right)B_l\rangle_B+
\langle B_l\left(t'\right)B_l^\dagger\rangle_B \right) \biggr. }}
\nonumber\ \\
&\times& \left[b^\dagger b \left(-t'\right) 
\rho^I\left( t \right),b^\dagger b \right] \nonumber \\
&+&\left( \langle B_l^\dagger\left(-t'\right)B_l \rangle_B+
\langle B_l\left(-t'\right) B_l^\dagger \rangle_B \right) \nonumber \\
&\times& \biggl. \left[b^\dagger b,
\rho^I\left( t \right) b^\dagger b \left(-t'\right)\right]
\biggr\}
\label{eqn:dampboson}
\end{eqnarray}
where 
\begin{eqnarray}
b^\dagger b \left( t \right)={\rm e}^{i H^I_{eff}t /\hbar}
b^\dagger b \/
{\rm e}^{-i H^I_{eff}t /\hbar}.
\end{eqnarray}

After expanding (\ref{eqn:dampboson}) in terms of the dressed states,
the time evolution of $\rho_{12}^{nn}$ and $\rho_{00}$ are
\begin{eqnarray}
\frac {\partial}{\partial t}\rho_{00}\left(t\right)&=&0,
\label{eqn:basiceq21}\\
\frac {\partial}{\partial t}\rho_{12}^{nn}\left(t\right)&=&
-i \Omega_n \rho_{12}^{nn}\left(t\right) \nonumber \\
&-& \frac{1}{2}
\left\{ \bar{n} \left( n \right)+1/2 \right\} \kappa \left(n \right)
\rho_{12}^{nn}\left(t\right) \nonumber \\ 
&+& \frac{1}{2}
\left\{ \bar{n} \left( n \right)+1/2 \right\} \kappa \left(n \right)
\rho_{21}^{nn}\left(t\right). 
\label{eqn:basiceq22}
\end{eqnarray}
The analytical solution of ${\rm Re} \left[ \rho_{12}^{nn} \right]$ is
given by (\ref{eqn:realpart}) with
\begin{eqnarray}
A_n= \frac{1}{2} \kappa \left( n \right)
\left \{ \bar{n} \left( n \right)+1/2 \right \} \equiv A_n^{vib}.
\label{eqn:damprate2}
\end{eqnarray}
$B_n$ is defined by the equation (\ref{eqn:coherentmotion}).  We note
that equations (\ref{eqn:basiceq21})-(\ref{eqn:basiceq22}) coincide
with those for the case of coupling to the reservoir via $S_z$
($H_{sr}=\hbar S_z \sum{g_{l}^\prime \left( B_l^\dagger+B_l\right)}$).

\section{Estimation of reservoir variables}

The formulation of the decoherence rates $A_n^{dipole}$
(\ref{eqn:damprate}) and $A_n^{vib}$ (\ref{eqn:damprate2}) in the
previous section shows that decoherence originates in the relaxation
of density matrix elements that are diagonal in the boson quantum
number but off-diagonal in the spin quantum numbers in the dressed
state basis.  The relaxation of the element $\rho_{\alpha \beta}^{nn}$
for $\alpha \neq \beta$ is caused by the coupling to reservoir bosons
at frequency of $\Omega_n$ ($=2 g \sqrt{n+1}$).  The effective
contribution of reservoir bosons at frequency of $\Omega_n$ is
therefore the key to understand the decoherence rate.

The Rabi frequency $g$ in the Boulder experiment \cite{Meekhof} is
around 100 kHz, so reservoir bosons of order $100$ kHz seem to be
responsible for decoherence.  These reservoir bosons have a much lower
frequency than those responsible for the case of spontaneous emission
between internal atomic states, which is of order GHz here, and also
population decay of motional states, which is of order 10 MHz.  This
low frequency nature of the reservoir bosons important here suggests
that the reservoir may be at non-zero temperature whereas of course in
the optical frequency regime, the reservoir is often approximated to
be at a zero temperature.

What are these reservoir bosons in the experiment?  To consider this
question and discuss the origin of decoherence, we investigate the
other characteristics, which the reservoir bosons should satisfy, by
comparing our theoretical results to the experiment.  For this
purpose, we introduce normalised values,
$\tilde{A}_n^{dipole}=A_n^{dipole}/g$,
$\tilde{A}_n^{vib}=A_n^{vib}/g$, $\tilde{\omega}_x=\omega_x/g$,
$\tilde{T}=k_B T/\hbar g$, $\tilde{\Omega}_n=\Omega_n/g$,
$\tilde{\gamma}_0=\gamma_0/g$, $\tilde{\kappa}\left(n \right)=\kappa
\left(n \right)/g$.  The normalised decoherence rates are
\begin{eqnarray} 
\tilde{A}_n^{dipole}&=&\left(n+1 \right) \tilde{\kappa} \left( n \right)
f \left( n,\tilde{T} \right )+2 \left( n+1 \right) \kappa_0 \bar{n}_0,\\
\tilde{A}_n^{vib}&=&\frac{1}{2} \tilde{\kappa} \left( n \right) 
f \left( n,\tilde{T}  \right)
\end{eqnarray}
where 
\begin{eqnarray}
f \left( n,\tilde{T} \right )= \bar{n} \left( n \right)+\frac{1}{2}
=\frac{1}{2}\coth \frac{\sqrt{n+1}}{\tilde{T}}.
\end{eqnarray}
The experimentally observed decoherence rate can be written as
$\tilde{A}_n^{ex}=\tilde{\gamma}_0 \left( n+1 \right)^\nu$.  In the
experiment, $g/2 \pi=94$ kHz, $\gamma_0=11.9$ kHz, so we have
$\tilde{\gamma}_0=0.127/2 \pi$.

Let us further assume that $\tilde{\kappa} \left( n \right)$ described
by (\ref{eqn:dampfunct}) is given by a power $d$ of the frequency
$\tilde{\Omega}_n$ \cite{Leggett},
\begin{equation}
\tilde{\kappa} \left( n \right) =\tilde{a} \tilde{\Omega}_n^d
=\tilde{a} \left(2 \sqrt{n+1} \right)^d
\label{eqn:rdpower}
\end{equation}
where $\tilde{a}$ is a damping constant ($\tilde{a} \ll 1$).  Some
high frequency cut-off of the damping function is assumed to prevent
divergence. These are the usual arguments given for the reservoir
density of states.  Generally, the case of $d=1$ is known as the Ohmic
case, since the choice of $d$ gives a velocity dependent dissipation
rate for the dissipative two-state system \cite{Leggett}, and $d=3$ is
required to describe 3-D radiation fields \cite{Keitel}.  However, we
do not restrict ourselves to $d$ as an integer.  We ignore the effect
of zero frequency reservoir bosons: $\kappa_0 \bar{n}_0=0$.

The decoherence rates at $n=0$ have to coincide with
$\tilde{\gamma}_0$.  So we have $\tilde{\gamma}_0=\tilde{\kappa} ( 0 )
f ( 0,\tilde{T} )$ for the imperfect dipole transition case and
$\tilde{\gamma}_0=\tilde{\kappa} ( 0 ) f (0,\tilde{T})/2$ for the case
of fluctuations of the vibrational potential. These conditions for
$\tilde{\gamma}_0$ determine $\tilde{a}$ when the value of
$\tilde{\gamma}_0$ is given.  Thus the decoherence rate is rewritten
as
\begin{eqnarray}
\tilde{A}_n^{dipole}&=&\tilde{\gamma}_0 \left( n+1 \right)^{1+d/2} 
f \left( n,\tilde{T} \right ) /f \left( 0,\tilde{T} \right ) , \\
\tilde{A}_n^{vib}&=&\tilde{\gamma}_0 \left( n+1 \right)^{d/2} 
f \left( n,\tilde{T} \right )/f \left( 0,\tilde{T} \right ).
\end{eqnarray}
The remaining unrestricted fitting parameters in our formulation are
the normalised temperature $\tilde{T}$ and the power dependency $d$ in
Eq.(\ref{eqn:bosonfunction}). The value $f \left( 0,\tilde{T}
\right )/ f \left( n,\tilde{T} \right )$ lies in the range
\begin{eqnarray}
\left(1+n \right)^{-1/2} \le 
 f \left( n,\tilde{T} \right )/f \left( 0,\tilde{T} \right ) \le 1.
\end{eqnarray}

\begin{figure}
  \begin{center}
    \leavevmode
    \epsfxsize=3.17truein
    \epsfbox{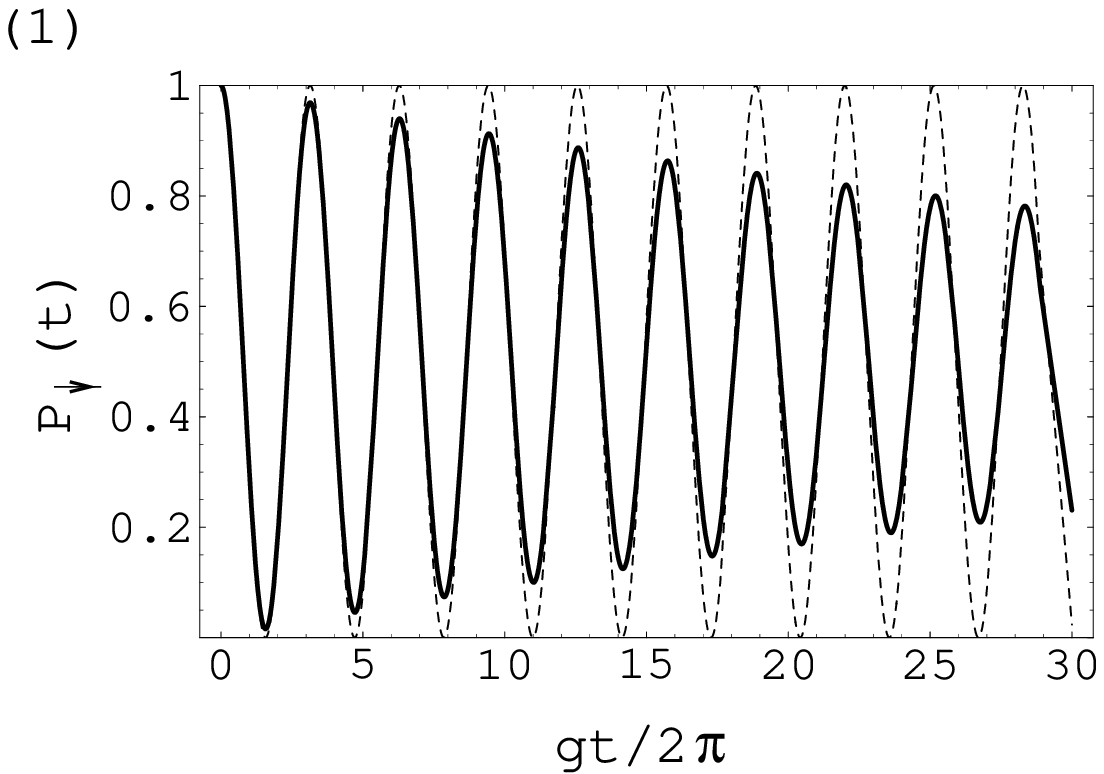}
    \epsfxsize=3.17truein
    \epsfbox{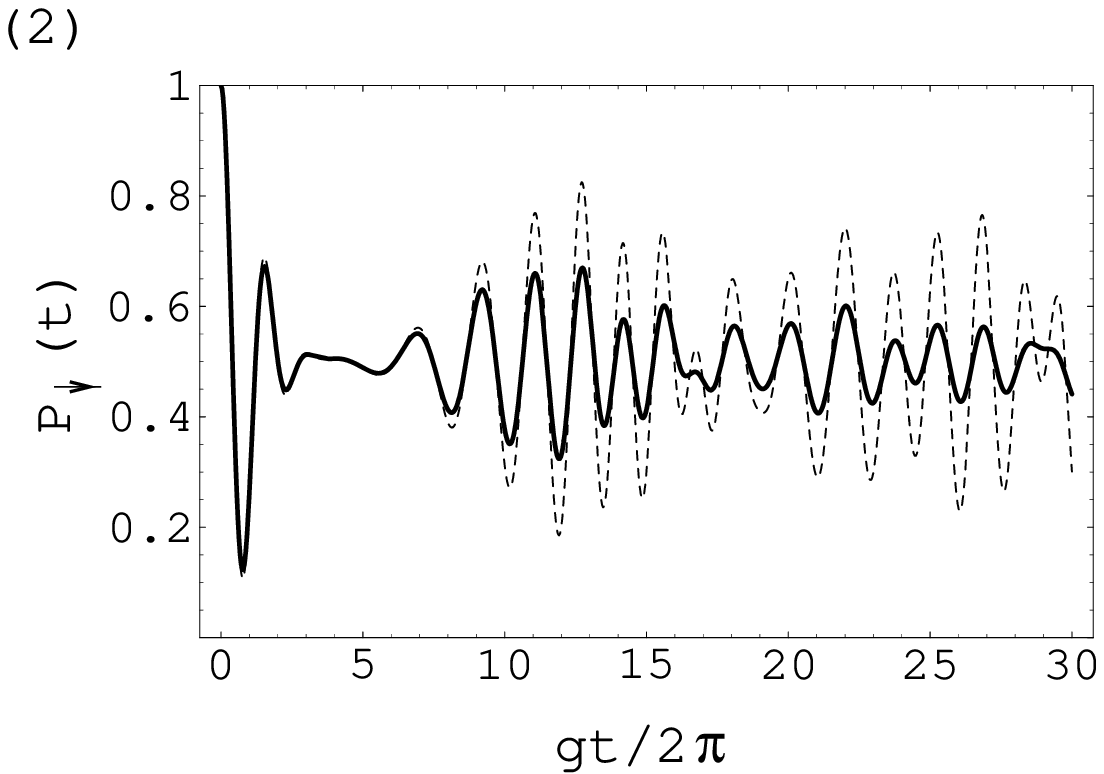}
  \end{center}
\caption{The population of the lower atomic state $P_\downarrow \left
( t \right)$ against the normalised time $gt/2\pi$ when (1) the
initial internal state is $\left \vert \downarrow \right \rangle$ and
the initial motional state is a {\it Fock} state with condition $\left
\vert 1 \right \rangle$; (2) the initial motional state is a {\it
coherent} state with condition $\left \vert \alpha=3.0 \right
\rangle$.  For both figures, the dashed lines are for the case of no
decoherence and the solid lines are for the case of an imperfect
dipole transition with the coefficients $d=0.4$ and ${\tilde
\gamma}_0=0.127/2 \pi$ in the high temperature limit $\tilde{T} \rightarrow
\infty$.}
\label{fig:fig}
\end{figure}

We take the high temperature limit ($\tilde{T} \rightarrow \infty$).
This limit represents classical noise where the reservoir operators
commute $\left( \left[B_l, B_l^\dagger \right]=0 \right)$. The value
$f \left( n,\tilde{T} \right )/f \left( 0,\tilde{T} \right )$
becomes $\left(n+1 \right)^{-1/2}$ when $\tilde{T} \rightarrow
\infty$, so we have
\begin{eqnarray}
\tilde{A}_n^{dipole}&=&\tilde{\gamma}_0 \left( n+1 \right)^{(d+1)/2}\\
\tilde{A}_n^{vib}&=&\tilde{\gamma}_0 \left( n+1\right)^{(d-1)/2}.
\end{eqnarray}
The linear form $\tilde{\gamma}_0 \left(n+1 \right)$ is reached for
$d=1$ (the Ohmic case) for the imperfect dipole transition and for
$d=3$ (3-D radiation field) for the case of fluctuations of the
vibrational potential.  To get a power exponent of $0.7$ for $n+1$ we
need $d \approx 0.4$ for the imperfect dipole transition case
(Fig.~\ref{fig:fig}) and $d \approx 2.4$ for the case of fluctuations
of the vibrational potential.

\end{multicols}
\smallskip
\begin{multicols}{2}

\section{Summary and open questions}

In summary, we have shown here a model describing decoherence which
destroys the characteristic quantum dynamics (collapse and revival) of
the Jaynes-Cummings system without energy relaxation for the ion trap
experiment \cite{Meekhof}.  The sources of decoherence are first
introduced phenomenologically and then described by a master equation
using a microscopic Hamilton formulation.  We apply the model to the
two possible actual sources of decoherence; one is the imperfect
dipole transition, and the other is the fluctuation of vibrational
potential. We solve the master equation under the Markovian
approximation and continuum limit of the reservoir modes.

The analytical solution shows that decoherence is described by the
reservoir bosons with frequency $\Omega_n =2 g \sqrt{n+1}$. Therefore,
the effective contribution of the bosons at frequency of $\Omega_n$
(which is of order 100 kHz) was found to be the key to understand the
decoherence. This low frequency nature of the reservoir bosons
compared to the spontaneous emission transition frequencies (which are
of order GHz) and population decay transition frequencies (which is of
order 10 MHz) suggests that the reservoir may be regarded to be at
non-zero temperature.  If we assume the high temperature limit and
certain density of states of the reservoir bosons, the decay rate
coincides with that seen in the experiment \cite{Meekhof}.

To proceed further and investigate the origin of decoherence for the
Boulder ion trap experiment \cite{Meekhof}, we would have to know a
number of parameters:
\begin{enumerate}
\item{The intensity fluctuations of the dye laser used for the
stimulated Raman transition which seems to be order of $10^5$-$10^6$
Hz, so this may well be a candidate for the fluctuation affecting the
bosons.  But we would need to know more about the frequency dependence
of the intensity fluctuation around 100 kHz to take the analysis much
further.}
\item{Noise from the trap potential from the radio frequency (100-200
kHz) radiation field is possible, but again we would need to estimate
the density of states for this case to be more precise.}
\item{Possibility of quantum noise (noise at finite $T$) remains a
potential candidate to explain this decoherence.}  
\end{enumerate}
We defer further consideration of all these until the underlying
parameters are better understood. Since this paper was submitted for
publication we learnt of related work by Schneider and Milburn
\cite{Schneider}, and by James \cite{James}.

\section*{Acknowledgement}

This work was supported in part by the Japan Society for the Promotion
of Science and the UK Engineering and Physical Sciences Research
Council, the European Community.  M.M. is grateful to P. Masiak,
J. Twamley and J. Steinbach for useful comments.


\end{multicols}


\begin{thebibliography}{99}
\bibitem{Meekhof} D.M. Meekhof, C. Monroe, B.E. King, W.M. Itano and
D.J. Wineland, Phys. Rev. Lett. {\bf 76}, 1796 (1996).
%
\bibitem{Shore} B.W. Shore and P.L. Knight, J. Mod. Opt. {\bf 40}
(1993) 1195.
%
\bibitem{Blockley} C.A. Blockley, D.F. Walls and H. Risken,
Europhys. Lett. {\bf 17} (1992) 509; W. Vogel and R.L. De Matos Filho,
Phys. Rev. A {\bf 49} (1995) 4214; J.I. Cirac, R. Blatt, A.S. Parkins
and P. Zoller, Phys. Rev. A {\bf 49} (1994) 1202.
%
\bibitem{Leibfreid} D. Leibfried, D.M. Meekhof, C. Monroe, B.E. King,
W.M. Itano and D.J. Wineland, J. Mod. Opt {\bf 44} (1997) 2485 and
references therein.
%
\bibitem{Wineland} D.J. Wineland, C. Monroe, W.M. Itano, D. Leibfried,
B. King and D.M. Meekhof, submitted to Rev. Mod. Phys. (1997).
%
\bibitem{Bose} S. Bose, P.L. Knight, M. Murao, M.B. Plenio and
V. Vedral, Phil. Trans. R. Soc. (in press), (1998).
%
\bibitem{Leggett} A.J. Leggett, S. Chakaravary, A.T. Dorsey,
M.P.A. Fisher, A. Garg and W. Zwerger, Rev. Mod. Phys. {\bf 59} (1987)
1.
%
\bibitem{Steinbach} J. Steinbach, J. Twamley, and P.L. Knight,
Phys. Rev. A {\bf 56} (1997) 4815.
%
\bibitem{Louisell} W.E. Louisell, {\it Quantum Statistical Properties
of Radiation} (John Wiley \& Sons, Inc., New York, 1973).
%
\bibitem{Zurek} W.H. Zurek, Phys. Today {\bf 44} (1991) 36.
%
\bibitem{Moya-Cessa} H. Moya-Cessa, V. Bu$\check{z}$ek, M.S. Kim and
P.L. Knight, Phys. Rev. A. {\bf 48} (1993) 3900. 
%
\bibitem{Milburn} G.J. Milburn, Phys. Rev. A {\bf 44} (1991) 5401. 
%
\bibitem{Shibata} F. Shibata and T. Arimitsu, J. Phys. Soc. Jpn {\bf 49} (1980)
891.
%
\bibitem{Murao2} M. Murao, J. Phys. Soc. Jpn. {\bf 66} (1997) 2314.
%
\bibitem{Murao3} M. Murao and F. Shibata, J. Phys. Soc. Jpn. {\bf 64}
(1995) 2394.
%
\bibitem{Keitel} C.H. Keitel, P.L.Knight, L.M. Marducci and
M.O. Scully, Opt. Commun. {\bf 118} (1995) 143.
%
\bibitem{Schneider}
S. Schneider and G.J. Milburn, quant-ph/9710044. 
%
\bibitem{James} D.F.V. James, {\it The thoery of heating of the
quantum ground state of trapped ions}, submitted for publication.
%
\end{thebibliography}
\end{document}